\documentclass[lineno]{jfm}

\usepackage{graphicx}
\usepackage{newtxtext}
\usepackage{newtxmath}
\usepackage{natbib}
\usepackage{hyperref}
\usepackage{color}

\hypersetup{
    colorlinks = false,
    urlcolor   = blue,
    citecolor  = black,
}

\newcommand{\RomanNumeralCaps}[1]
\linenumbers

\title{Predicting drag on rough surfaces by transfer learning of empirical correlations}

\author{Sangseung Lee\aff{1}
  \corresp{\email{sanlee@mech.kth.se}},
  Jiasheng Yang\aff{2},
  Pourya Forooghi\aff{3},
  Alexander Stroh\aff{2},
 \and Shervin Bagheri\aff{1}
   \corresp{\email{shervin@mech.kth.se}}}

\affiliation{\aff{1}Department of Engineering Mechanics, FLOW Centre, KTH, Stockholm SE-100 44, Sweden
\aff{2}Institute of Fluid Mechanics, Karlsruhe Institute of Technology, Karlsruhe 76131, Germany
\aff{3}Department of Mechanical and Production Engineering, Aarhus University, 8000, Aarhus C, Denmark}

\begin{document}
\maketitle

\begin{abstract}
Recent developments in neural networks have shown the potential of estimating drag on irregular rough surfaces. Nevertheless, the difficulty of obtaining a large high-fidelity dataset to train neural networks is deterring their use in practical applications. In this study, we propose a transfer learning framework to model the drag on irregular rough surfaces even with a limited amount of direct numerical simulations. We show that transfer learning of empirical correlations, reported in the literature, can significantly improve the performance of neural networks for drag prediction. This is because empirical correlations include `approximate knowledge' of the drag dependency in high-fidelity physics. The `approximate knowledge' allows neural networks to learn the surface statistics known to affect drag more efficiently. The developed framework can be applied to applications where acquiring a large dataset is difficult, but empirical correlations have been reported.
\end{abstract}


\section{Introduction}
Nearly all surfaces in fluid-related industries are rough at their operating Reynolds numbers. These rough surfaces easily alter the aero- or hydrodynamic properties of a fluid system and induce unfavorable consequences; for instance, reduced stability of aircraft, increased fuel consumption of cargo ships, and reduced energy harvesting capacity of wind turbines~\citep{gent2000aircraft,schultz2011economic,dalili2009review}. Therefore, it is crucial to restore or replace surfaces that have degraded past a critical level. This requires monitoring roughness on industrial surfaces and efficiently predicting their effect on the flow. 

One of the most important effects of surface roughness on flow is the increased drag. This additional drag is generally expressed by the Hama roughness function~\citep{hama1954boundary} or the equivalent sand-grain height. The Hama roughness function is the downward shift of the mean velocity profile  induced by roughness. The equivalent sand-grain height is the roughness height from the Moody diagram~\citep{moody1944friction} that would cause the same drag as the rough surface of interest. It is worth noting that to calculate the equivalent sand-grain height, the Hama roughness function in the fully-rough regime -- where the skin-friction coefficient is independent of Reynolds number -- needs to be determined. 
To accurately determine the Hama roughness function of irregular surfaces, numerical simulations with fully resolved rough structures or experiments are needed. However, neither simulations nor experiments are cost-efficient, hence not feasible for practical purposes. To alleviate the high costs, researchers have found empirical correlations between rough surface statistics and their contribution to drag~\citep{chan2015systematic,forooghi2017toward,flack2010review,flack2020skin}. However, no universal model has emerged that is able to accurately predict the drag from a sufficiently wide range of roughness. This is mainly attributed to the high-dimensional feature space of irregular rough structures~\citep {chung2021predicting}.

Neural networks are known to be capable of extracting patterns in high-dimensional spaces, and therefore have been successfully used in various studies using fluid flow data~\citep{lee2019data,kim2020prediction,fukami2021machine}. \citet{jouybari2021data} developed a multilayer perceptron (MLP) type neural network to find a mapping of 17 different rough surface statistics to equivalent sand-grain height. They reported a state-of-the-art performance in predicting equivalent sand-grain heights from rough surface statistics. A total of 45 data samples, obtained from direct numerical simulations (DNS) and experiments, were used for training and validating their neural network. Having a large number of data samples to train a neural network is invariably advantageous; nonetheless, constructing a fluid database that is sufficiently large to train a network from scratch imposes significant computational or experimental cost and is oftentimes considered impractical. Therefore, for the practical usage of neural networks in fluid-related applications, a framework that allows neural networks to learn a generalized mapping from a limited number of data samples is needed.

In this study, we propose a transfer learning framework to improve the generalization ability of neural networks for predicting the drag on rough surfaces when only a limited number of high-fidelity data samples are available. Transfer learning is a method that improves learning performance by adapting knowledge learned from one domain to another~\citep{pan2009survey,chakraborty2021transfer}. 
The method has been used, for example, for self-driving cars, which can be pre-trained with data from virtual car simulators~\citep{pan2017virtual,martinez2017beyond,akhauri2020enhanced}. Obviously, the driving environment in virtual car simulators is inherently different from the real world, however, pre-training a neural network with `approximate knowledge' can provide better initial weights for training real self-driving cars.

Similarly, for flow over rough surfaces, empirical correlations between drag and surface roughness provide an approximate knowledge of real flow physics~\citep{colebrook1937experiments,moody1944friction,chan2015systematic,forooghi2017toward,flack2010review,flack2020skin}. These correlations were developed by a fitting procedure of large experimental and numerical datasets. However, it is not straightforward to make direct use of these datasets, in particular because the information of surface topographies and flow statistics is not always accessible. Indeed, in many cases we have advanced our understanding of physics from the valuable information embedded in empirical correlations, such as those developed by~\cite{colebrook1937experiments} and~\citet{moody1944friction}.

The aim of this study is to show that transfer learning of empirical correlations can significantly improve the performance of neural networks for modeling drag on rough surfaces. Our aim is also to analyze a simple neural network with and without transfer learning to gain insight of the learning behavior, which is strongly connected to the physics of the problem. The objective of the developed neural network model is to predict drag of a particular class of roughness contained in one high-fidelty dataset, which is composed of irregular homogeneous rough surfaces. However, we foresee that transfer learning will play a central role in combining empirical relations with larger datasets from a range of sources to  develop  predictive models of a significantly larger class of rough walls than the ones considered here. The paper is organized as follows: the details of the developed transfer learning framework including the neural networks and datasets are explained in \S\ref{sec:TF}. The results of learning a mapping of surface statistics to the Hama roughness function using the transfer learning framework are shown in~\S\ref{sec:Result}. The effects of transfer learning are analyzed and discussed in~\S\ref{sec:discussion}, followed by the concluding remarks in~\S\ref{sec:conclusion}.
\begin{figure}
  \centerline{\includegraphics[width = 1.0 \linewidth]{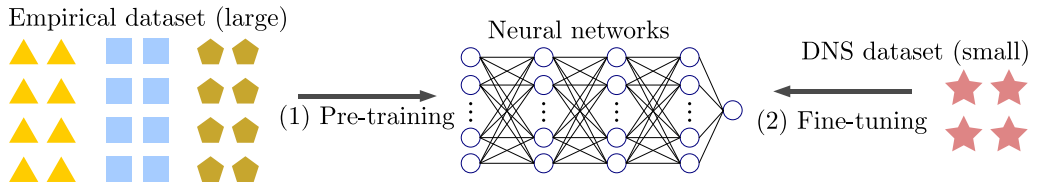}}
  \caption{Overview of the transfer learning framework.}
\label{fig:tf_overview}
\end{figure}

\section{Transfer learning framework}\label{sec:TF}
The developed transfer learning framework is composed of two parts: (1) pre-training step and (2) fine-tuning step (figure~\ref{fig:tf_overview}). In the pre-training step, neural networks were trained to learn an `approximate’ knowledge of the mapping of surface statistics to the Hama roughness function. The training data is created by evaluating empirical correlation functions given the surface statistics of different surface topographies. In the fine-tuning step, neural networks were tuned to learn high-fidelity physics from a small DNS dataset by adapting their pre-trained `approximate’ knowledge to the domain of real physics. In \S\ref{sec:Result}, we show that transfer learning of empirical correlations improves the generalization ability of neural networks when learning the drag of rough surfaces from a small amount of data. First, however, we explain the details of the neural networks and pre-training and fine-tuning steps.

\subsection{Neural network}
We used a variation of the MLP type neural network architecture developed by~\citet{jouybari2021data}. The input of the employed network is a set of 17 different rough surface statistics calculated from the surface topography of interest. The 17 input parameters are composed of 8 primary surface statistics and 9 products of the 8 primary statistics. These parameters are selected based on their perceived importance on affecting drag on rough surfaces~\citep{jouybari2021data}. 

Let $x$, $y$, and $z$ be the streamwise, wall-normal and spanwise directions, $k(x,z)$ the roughness height distribution and $A_t$ the total roughness plan area. Then, one may define the following measures of an irregular rough surface: 
\begin{equation}
    \begin{array}{llcl}
   \textrm{average roughness height: }&k_{avg}&=&\int_{x,z}k dA / A_{t},\\
\textrm{average height deviation: }&Ra  &=&\int_{x,z}(|k-k_{avg}|) dA / A_{t},\\
\textrm{root-mean-squared roughness height: } & k_{rms} & =&\left (\int_{x,z}(k-k_{avg})^{2}dA/A_{t}\right )^{1/2}.
    \end{array}
\end{equation}
In addition, we define the crest height $k_{c}$, and finally, the fluid area at each wall-normal plane $A_{f}(y)$. From these quantities, the 17 statistical input parameters ($\{I_{1}, I_{2}, ..., I_{17}\}$) can be calculated as listed in table~\ref{tab:I}.
\begin{table}
  \begin{center}
\def~{\hphantom{0}}
  \begin{tabular}{ll}
      $I_{1} = k_{rms}/Ra$  & $I_{2} = Sk = \frac{1}{A_{t}}\int_{x,z}(k-k_{avg})^{3} dA / k^{3}_{rms}$ \\
      $I_{3} = kur = \frac{1}{A_{t}}\int_{x,z}(k-k_{avg})^{4} dA / k^{4}_{rms}$ & $I_{4} = ES_{x} = \frac{1}{A_{t}}\int_{x,z}|\frac{\partial k}{\partial x}| dA$ \\
      $I_{5} = ES_{z} = \frac{1}{A_{t}}\int_{x,z}|\frac{\partial k}{\partial z}| dA$  & $I_{6} = por = \frac{1}{A_{t}k_{c}}\int_{0}^{k_{c}} A_{f} dy$ \\
      $I_{7} = inc_{x} = \tan^{-1}\{\frac{1}{2}Sk(\frac{\partial k}{\partial x})\}$   & $I_{8} = |inc_{z}| = |\tan^{-1}\{\frac{1}{2}Sk(\frac{\partial k}{\partial z})\}|$ \\
      $I_{9} = I_{4}I_{5}$  & $I_{10} = I_{4}I_{2}$ \\
      $I_{11} = I_{4}I_{3}$ & $I_{12} = I_{5}I_{2}$ \\
      $I_{13} = I_{5}I_{3}$ & $I_{14} = I_{2}I_{3}$ \\
      $I_{15} = I_{4}I_{4}$ & $I_{16} = I_{5}I_{5}$ \\
      $I_{17} = I_{2}I_{2}$ & 
  \end{tabular}
  \caption{Statistical input parameters of the employed neural network.}
  \label{tab:I}
  \end{center}
\end{table}
Here, $Sk$, $kur$, $por$, $ES_{x}$, $ES_{z}$, $inc_{x}$, and $inc_{z}$ indicate skewness, kurtosis, porosity, and effective slopes and average inclinations in $x$ and $z$ directions, respectively. Before training, the parameters $I_{1}$ to $I_{17}$ were normalized with the mean and standard deviation values calculated from the dataset of empirical correlations (see section~\ref{sec:pre-train} for the details of the empirical dataset).

\begin{figure}
  \centerline{\includegraphics[width = 0.9 \linewidth]{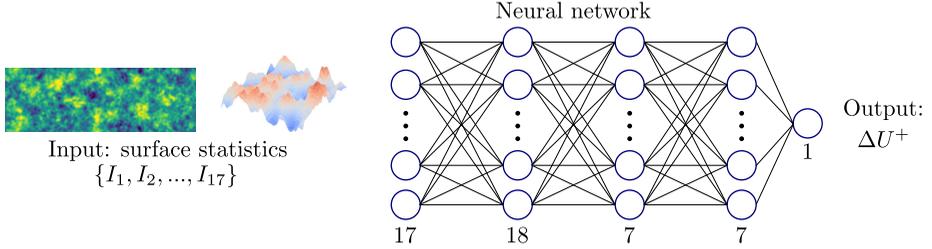}}
  \caption{Architecture of the used neural network.}
\label{fig:NN}
\end{figure}

Information of input parameters travels through three hidden layers with 18, 7, and 7 neurons, respectively (see figure~\ref{fig:NN}). Leaky rectified linear unit (ReLU) activations ($\max(0.01x,x)$) were applied after the hidden layers. The output of the employed neural network is a scalar value of the Hama roughness function $\Delta U^{+}$. The superscript $+$ indicates the normalization by the viscous scale $\delta_{\nu}=\nu/u_{\tau}$, where $\nu$ and $u_{\tau}$ are the kinematic viscosity and the friction velocity, respectively. In this study, the friction Reynolds number $Re_{\tau}=u_{\tau} H/ \nu=500$ was used, where $H$ is the channel half-height. At the output layer, a Sigmoid activation function, $a/(1+e^{x})$,  was applied to bound the prediction value ($a=20$ in this study).

\subsection{Pre-training step}\label{sec:pre-train}
In the pre-training step, neural networks were trained with a large dataset obtained from empirical correlation functions between rough surface statistics and the Hama roughness function $\Delta U^{+}$. We used three correlations; 
\begin{equation}
  \Delta U^{+} = 2.5 \log Ra^{+} + 1.12 \log ES_{x} + 1.47;\label{EMP1}\tag{EMP1}
\end{equation}
\begin{equation}
    \Delta U^{+}= 2.5 \log \left [k^{+}_{c}\left (0.67 Sk^{2}+0.93 Sk + 1.3\right)\left (1.07\left (1-e^{-3.5 ES_{x}}\right )\right)\right ]- 3.5; \label{EMP2}\tag{EMP2}
\end{equation}
and
\begin{equation}
  \Delta U^{+} = \left\{
    \begin{array}{ll}
      2.5 \log \left [2.48 k^{+}_{rms}(1+Sk)^{2.24}\right ] - 3.5, & Sk > 0 \\
            2.5 \log \left [2.11 k^{+}_{rms}\right ] - 3.5, & Sk \approx 0, \\
      2.5 \log \left [2.73 k^{+}_{rms}(2+Sk)^{-0.45}\right ] - 3.5, & Sk < 0.
    \end{array} \right. \label{EMP3}\tag{EMP3}
\end{equation}
Here, \ref{EMP1} was proposed by \citet{chan2015systematic}, \ref{EMP2} by \citet{forooghi2017toward} and \ref{EMP3} by \citet{flack2010review,flack2020skin}. 

As mentioned earlier, we will develop neural networks to predict drag of rough surfaces contained in a small high-fidelity dataset (see \S\ref{sec:finetune} for the details of the DNS dataset). The range of the roughness function in this dataset is $5.3<\Delta U^{+}<8.3$. The lower bound lies at the boundary between transitionally and fully rough regimes. \ref{EMP1} was derived using surfaces in the transitionally and fully rough regimes with $0.4<\Delta U^{+}<11.4$. \ref{EMP2} was modeled from surfaces with roughness function in the range of $5.4<\Delta U^{+}<10.1$. Finally, \ref{EMP3} was developed with rough surfaces in the fully rough regime, where most of the surfaces become fully rough when $\Delta U^{+}$ was over $5.5-6.0$.  Therefore, all three empirical models contain `approximate' knowledge of the drag of rough surfaces that can help the neural networks to learn the DNS dataset in the fine-tuning step.

\begin{figure}
  \centerline{\includegraphics[width = 1.0\linewidth]{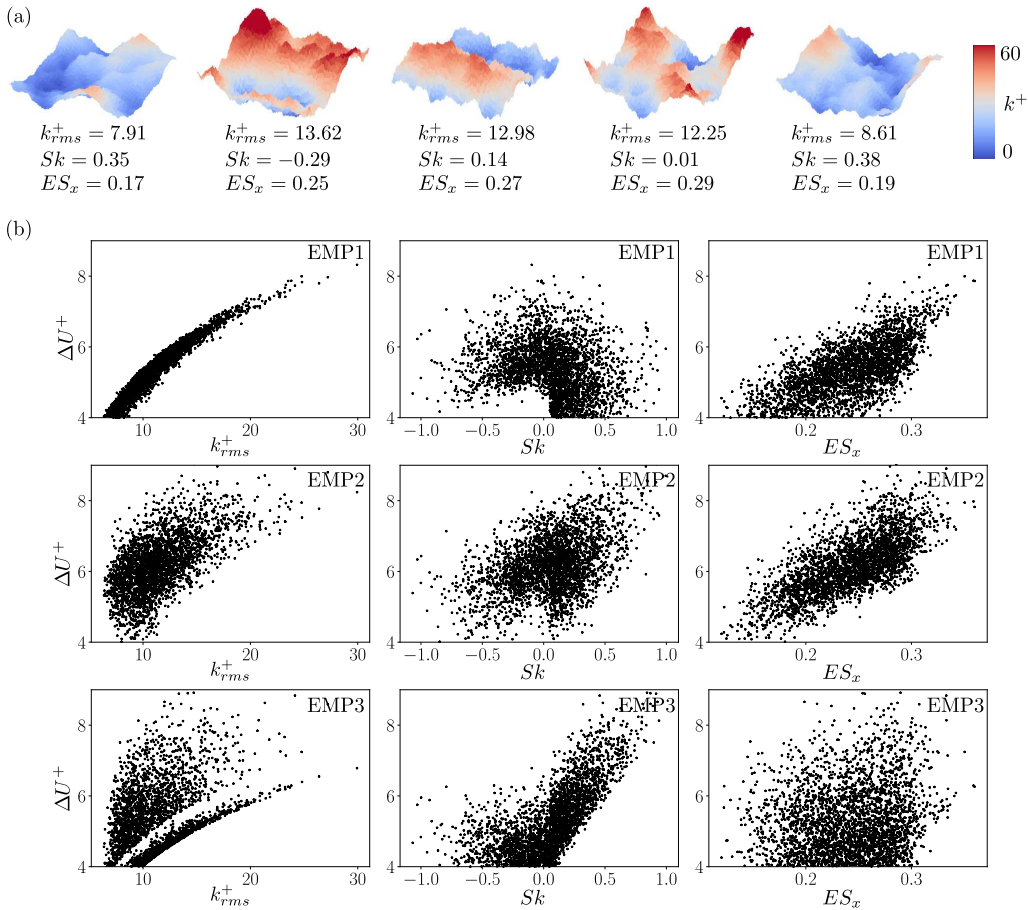}}%
  \caption{(a) Examples of patches of the generated surfaces in the empirical dataset. The size of the patches is $500 \times 500$ (viscous unit) in the streamwise and spanwise directions. The visualized heights of the surfaces are exaggerated five times in order to show the topography explicitly. (b) Scatter plots of the Hama roughness functions ($\Delta U^{+}$), calculated by \ref{EMP1}, \ref{EMP2}, and \ref{EMP3}, against $k_{rms}^{+}$, $Sk$, and $ES_{x}$ from the 3,000 validation data samples.}
\label{fig:emp}
\end{figure}

A large dataset was constructed by randomly generating 10,000 irregular rough surfaces and calculating the associated Hama roughness function using \ref{EMP1}, \ref{EMP2} and \ref{EMP3}. The surfaces were constructed using the Fourier-filtering algorithm suggested by \citet{jacobs2017quantitative}. The algorithm generates self-affine surfaces with a pre-defined power spectrum $C$,
\begin{equation}
 C(q) \propto q^{-2-2h}
\label{eq:self-affine}   
\end{equation}
where $q$ is the wavenumber and $h$ the Hurst exponent. The Hurst exponent $h=0.8$ was used as in \citet{jacobs2017quantitative}. This power-law dependence is a well-known attribute of self-affine realistic surfaces~\citep{mandelbrot1982fractal,persson2004nature}. 
Figure~\ref{fig:emp}a shows a few examples of the generated random surfaces. The randomness is imposed by choosing random amplitudes of the power spectrum and random phase shifts of Fourier modes. Accordingly, the surface statistics, such as the skewness and the effective slope, depend on the particular combination of the random amplitudes and phases of Fourier modes.

For training and validating the neural network, 7,000 and 3,000 data samples were used, respectively. Figure~\ref{fig:emp}b shows the scatter plots of the Hama roughness functions, calculated by \ref{EMP1}, \ref{EMP2} and \ref{EMP3}, against $k_{rms}^{+}$, $Sk$, and $ES_{x}$ from the 3,000 validation data samples.
The surface generation method based on the spectral density \eqref{eq:self-affine}, has generated surfaces with skewness bounded between $-1$ to $+1$ and effective slopes limited to $\lesssim 0.35$.
The bounded skewness and limited effective slopes are due to the wavy surfaces generated by Fourier modes and the self-affine power spectrum defined for realistic surfaces, as in~\citet{jacobs2017quantitative}. 
The random surfaces can thus be considered to be in the low-slope regime, where the drag tends to increase with increasing effective slope~\citep{flack2014roughness}. 
From figure~\ref{fig:emp}b, we observe different distributions, or knowledge, of the Hama roughness function from the different empirical models. The aim is to seed this diverse knowledge of empirical correlations in neural networks during the pre-training step.

Three neural networks with the same architecture and the same initial weights and biases were trained simultaneously to learn the different empirical models, \ref{EMP1} to \ref{EMP3}. Let $NN^{i}$ be the neural network trained with the $i$-th empirical model, $W^{i}_{j,k}$ the weight matrix connecting the $j$-th and $k$-th layers of $NN^{i}$, $N_{w}$ the total number of weights in $NN^{i}$, $b^{i}_{j}$ the bias vector added in the $j$-th layer, $\Delta U^{+}_{i}$ the Hama roughness function predicted by the $i$-th empirical model, and $\Delta \widetilde{U}^{+}_{i}$ the Hama roughness function predicted by $NN^{i}$. The neural networks were trained to minimize a loss for pre-training:
\begin{equation}
    L_{p} = \frac{1}{3}\sum_{i=1}^{3} \left[ \left(\Delta U^{+}_{i} - \Delta \widetilde{U}^{+}_{i} \right)^{2} + \sum_{j,k} \mathrm{tr}\left\{(W^{i}_{j,k})^{T}(W^{i}_{j,k})\right\} \right].
\end{equation}
The first term in the right-hand side is the mean-squared-error loss and the second is the weight regularization loss, also used in \citet{jouybari2021data}. The sizes of the weight matrix $W^{i}_{j,k}$ and bias vector $b^{i}_{k}$ are $n_{j} \times n_{k}$ and $n_{k}$, respectively, where $n_{j}$ and $n_{k}$ are the number of neurons on the $j$-th and $k$-th layers. 

The weights and biases of the neural networks were updated in the direction of minimizing the pre-training loss $L_{p}$ on the training data samples through the Adam optimizer~\citep{kingma2014adam}. The pre-training loss $L_{p}$ on the validation data samples was also evaluated each iteration of updates. Batch sizes of 16 and 32 were used for training and validating, respectively. To provide a wide distribution of data samples in the batches, i.e., preventing neural networks from learning biased data focused on average characteristics, we imposed batches to contain 50\% of the data samples with $\Delta U^{+}$ in the range 6 to 8, 25\% of the data samples with $\Delta U^{+}$ in the range 4 to 6, and the remaining 25\% of the data samples with $\Delta U^{+}$ in the range 8 to 10.
After a sufficient amount of iterations for training and validating, the set of networks ($\{NN^{1},NN^{2},NN^{3}\}$) that produced the lowest pre-training loss on the validation data samples was chosen to be merged into a single pre-trained network $NN^{pre}$. The learned sets of weights and biases in the three neural networks, $\{W^{1}_{j,k},W^{2}_{j,k},W^{3}_{j,k}\}$ and $\{b^{1}_{k},b^{2}_{k},b^{3}_{k}\}$, were averaged into the weights and biases of $W^{pre}_{j,k}$ and $b^{pre}_{k}$ as $W^{pre}_{j,k} = \sum_{i=1}^{3} W^{i}_{j,k}/3$ and $b^{pre}_{k} = \sum_{i=1}^{3} b^{i}_{k}/3$.
These pre-trained weights and biases were employed as the initial weights for the fine-tuning step (section~\ref{sec:finetune}) to transfer the knowledge of empirical correlations to the domain of high-fidelity physics. 

A key factor for the performance of the fine-tuned neural network is the contents of database used in the pre-training step. If the bound of the $\Delta U^{+}$ in one's high fidelity dataset is known \textit{a priori}, then by using a narrow range of $\Delta U^{+}$ for pre-training increases the transfer learning performance significantly. We provide an example of this by training the neural networks with a bound of $\Delta U^{+}$ in Appendix~\ref{appBU}. In the main text, we assume that the range of $\Delta U^{+}$ of the high-fidelity dataset is \textit{a priori} unknown.

\subsection{Fine-tuning step}\label{sec:finetune}
\begin{figure}
  \centerline{\includegraphics[width = 1.0 \linewidth]{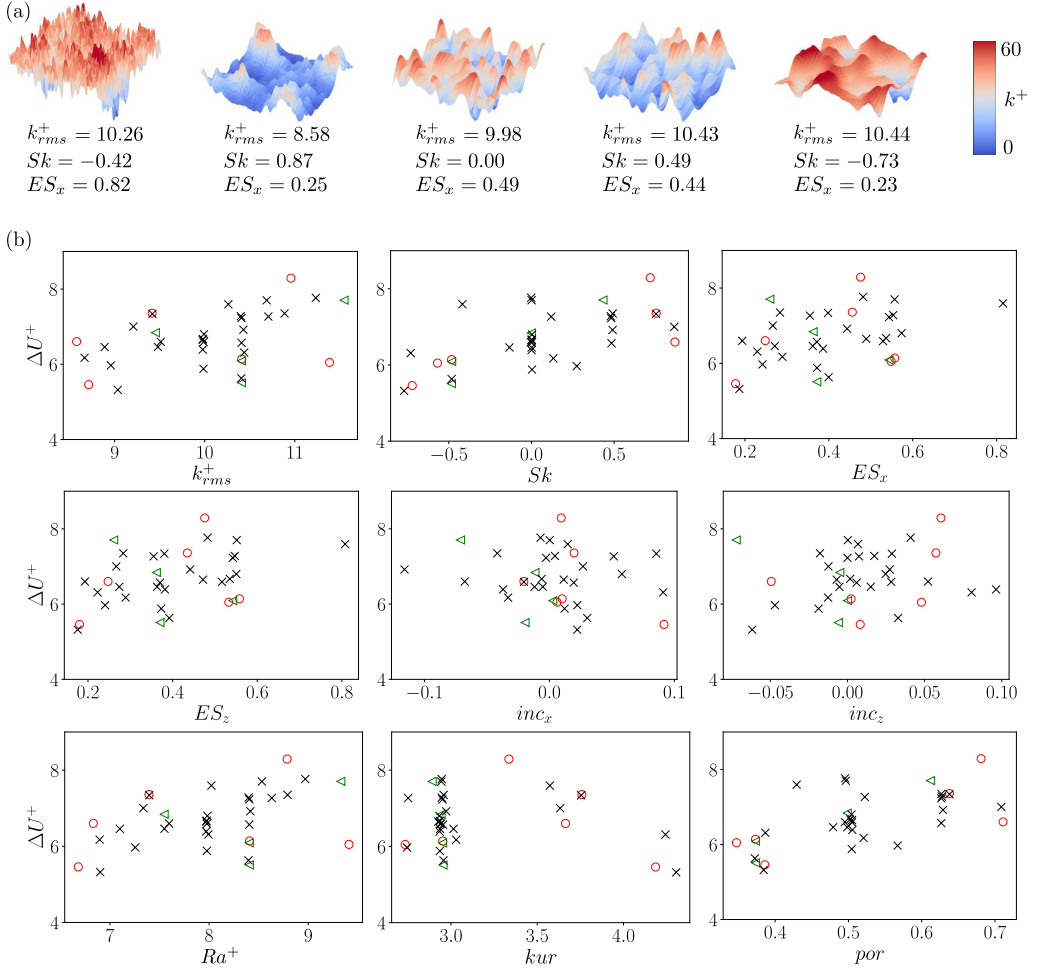}}
  \caption{(a) Examples of patches of the generated surfaces in the DNS dataset. The size of the patches is $500 \times 500$ (viscous unit) in the streamwise and spanwise directions. The visualized heights of the surfaces are exaggerated five times in order to show the topography explicitly. (b) Scatter plots of the calculated Hama roughness function ($\Delta U^{+}$) against various surface statistics of the 35 DNS data samples. The markers {\color{red}$\circ$}, {\color{green}$\lhd$}, and $\times$ indicate the training, validation, and test data samples, respectively. The plots were drawn from an example of a combination of training and validation data samples.}
\label{fig:dns}
\end{figure}

A total of 35 DNS of channel flow over rough surfaces at $Re_{\tau}\approx500$ were conducted to construct a high-fidelity dataset. The 35 rough surfaces were generated following the method proposed by \citet{perez2019generating}. Figure~\ref{fig:dns}a shows a few examples of the generated surfaces. Note that different surface generation methods were used to construct the DNS and empirical datasets (section~\ref{sec:pre-train}). The method used for DNS imposes an additional constraint on the roughness height probability distribution. This constraint enables the generation of rough surfaces with distinctive characteristics. For the probability distribution, we used Weibull distributions of the form
\begin{equation}
 f(x;\lambda,s) = \frac{s}{\lambda}\left(\frac{x}{\lambda}\right)^{s-1}e^{-(x/\lambda)^{s}}.   
 \label{eq:Weibull}
\end{equation}
By using different shape ($s$) and scale ($\lambda$) parameters, rough surfaces with non-zero skewness are generated. Gaussian distributions were used to generate rough surfaces with zero skewness. Rough surfaces generated with \eqref{eq:Weibull} imitate the roughness caused by wear~\citep{perez2019generating}. By using different surface generation methods in the pre-training and fine-tuning steps, we can verify that pre-training on one class of rough surfaces can help learn the drag on a different class of rough surfaces. Such verification is important, as in practice, artificially generated rough surfaces for pre-training would not perfectly imitate real-world rough surfaces.

\begin{table}
  \begin{center}
\def~{\hphantom{0}}
\begin{tabular}{lllllllllll}
Fine-tuning data    & $\Delta U^{+}$ & $k_{rms}^{+}$   & $Sk$    & $ES_{x}$    & $ES_{z}$    & $inc_{x}$     & $inc_{z}$     & $Ra^{+}$    & $kur$   & $por$   \\
F1  & 5.46  & 8.712  & -0.721 & 0.178 & 0.180 & 0.092  & 0.008  & 6.681 & 4.186 & 0.386 \\
F2  & 5.51  & 10.408 & -0.484 & 0.371 & 0.370 & -0.019 & -0.006 & 8.406 & 2.954 & 0.373 \\
F3  & 6.05  & 11.385 & -0.568 & 0.549 & 0.533 & 0.006  & 0.048  & 9.411 & 2.734 & 0.347 \\
F4  & 6.09  & 10.407 & -0.483 & 0.543 & 0.542 & 0.003  & 0.000  & 8.405 & 2.951 & 0.373 \\
F5  & 6.14  & 10.411 & -0.483 & 0.557 & 0.558 & 0.010  & 0.002  & 8.407 & 2.952 & 0.373 \\
F6  & 6.60  & 8.578  & 0.868  & 0.248 & 0.247 & -0.020 & -0.049 & 6.832 & 3.664 & 0.711 \\
F7  & 6.84  & 9.453  & -0.001 & 0.363 & 0.361 & -0.012 & -0.005 & 7.547 & 2.943 & 0.498 \\
F8  & 7.36  & 9.417  & 0.755  & 0.456 & 0.435 & 0.020  & 0.057  & 7.394 & 3.759 & 0.638 \\
F9  & 7.70  & 11.546 & 0.433  & 0.260 & 0.261 & -0.071 & -0.072 & 9.328 & 2.890 & 0.612 \\
F10 & 8.29  & 10.956 & 0.718  & 0.476 & 0.476 & 0.009  & 0.061  & 8.786 & 3.336 & 0.681 \\ \hline \hline
Test data & $\Delta U^{+}$ & $k_{rms}^{+}$   & $Sk$    & $ES_{x}$    & $ES_{z}$    & $inc_{x}$     & $inc_{z}$     & $Ra^{+}$    & $kur$   & $por$   \\
T1        & 5.32 & 9.035  & -0.770 & 0.187 & 0.176 & 0.022  & -0.062 & 6.902 & 4.306 & 0.384 \\
T2        & 5.63 & 10.406 & -0.483 & 0.399 & 0.392 & 0.030  & 0.033  & 8.397 & 2.956 & 0.372 \\
T3        & 5.88 & 9.986  & 0.003  & 0.372 & 0.373 & 0.012  & -0.019 & 7.978 & 2.936 & 0.504 \\
T4        & 5.97 & 8.958  & 0.274  & 0.242 & 0.240 & 0.022  & -0.047 & 7.256 & 2.745 & 0.567 \\
T5        & 6.17 & 8.667  & 0.132  & 0.289 & 0.289 & -0.033 & -0.013 & 6.897 & 3.030 & 0.521 \\
T6        & 6.31 & 10.439 & -0.730 & 0.230 & 0.223 & 0.091  & 0.080  & 7.993 & 4.243 & 0.387 \\
T7        & 6.39 & 9.978  & 0.001  & 0.387 & 0.382 & -0.037 & 0.096  & 7.972 & 2.934 & 0.505 \\
T8        & 6.46 & 8.889  & 0.001  & 0.271 & 0.274 & -0.012 & -0.005 & 7.100 & 2.937 & 0.498 \\
T9        & 6.46 & 9.481  & -0.133 & 0.362 & 0.363 & -0.005 & 0.015  & 7.550 & 3.015 & 0.479 \\
T10       & 6.57 & 10.408 & 0.484  & 0.371 & 0.370 & 0.019  & 0.006  & 8.406 & 2.954 & 0.627 \\
T11       & 6.59 & 9.986  & 0.003  & 0.529 & 0.517 & -0.020 & 0.028  & 7.977 & 2.937 & 0.505 \\
T12       & 6.60 & 9.512  & -0.004 & 0.194 & 0.193 & -0.068 & 0.052  & 7.597 & 2.943 & 0.495 \\
T13       & 6.65 & 9.976  & 0.001  & 0.489 & 0.472 & 0.011  & -0.007 & 7.972 & 2.926 & 0.505 \\
T14       & 6.67 & 9.980  & 0.002  & 0.537 & 0.535 & -0.006 & 0.002  & 7.973 & 2.936 & 0.502 \\
T15       & 6.80 & 9.992  & 0.003  & 0.574 & 0.551 & 0.058  & 0.025  & 7.983 & 2.933 & 0.504 \\
T16       & 6.92 & 10.429 & 0.491  & 0.444 & 0.441 & -0.116 & 0.027  & 8.416 & 2.972 & 0.629 \\
T17       & 7.00 & 9.208  & 0.864  & 0.266 & 0.267 & 0.027  & -0.012 & 7.337 & 3.633 & 0.709 \\
T18       & 7.23 & 10.407 & 0.483  & 0.543 & 0.542 & -0.003 & 0.000  & 8.405 & 2.951 & 0.627 \\
T19       & 7.27 & 10.706 & 0.120  & 0.355 & 0.355 & 0.051  & 0.007  & 8.632 & 2.751 & 0.522 \\
T20       & 7.28 & 10.400 & 0.480  & 0.554 & 0.547 & 0.004  & 0.017  & 8.399 & 2.946 & 0.627 \\
T21       & 7.34 & 9.419  & 0.754  & 0.398 & 0.380 & 0.086  & 0.029  & 7.397 & 3.755 & 0.638 \\
T22       & 7.35 & 10.886 & 0.491  & 0.284 & 0.283 & -0.042 & -0.018 & 8.792 & 2.957 & 0.627 \\
T23       & 7.59 & 10.262 & -0.418 & 0.815 & 0.808 & 0.015  & 0.007  & 8.023 & 3.573 & 0.429 \\
T24       & 7.70 & 10.688 & 0.001  & 0.557 & 0.552 & 0.000  & 0.000  & 8.532 & 2.944 & 0.496 \\
T25       & 7.77 & 11.233 & -0.001 & 0.482 & 0.483 & -0.007 & 0.041  & 8.970 & 2.946 & 0.495
\end{tabular}
  \caption{The data of roughness function ($\Delta U^{+}$) and surface statistics used for fine-tuning and testing the neural networks.}
  \label{tab:data}
  \end{center}
\end{table}

Details of the methodology and validation of the conducted DNS are reported in the study of~\citet{jiasheng2021dns}. Here, we provide a brief overview.
A pseudo-spectral incompressible Navier–Stokes solver, SIMSON~\citep{chevalier2007simson} with an immersed boundary method~\citep{goldstein1993modeling}, was employed for the channel simulations. Periodic boundary conditions were used in the streamwise and spanwise directions. No-slip conditions were applied at the top and bottom planes of the channel. Roughness effects were added to both the top and bottom planes. A minimal channel approach was adopted to find a small computational domain size that can accurately reproduce the Hama roughness function calculated from a full channel simulation~\citep{chung2015fast}.
Further details on the domain size and grid resolution are provided in Appendix~\ref{appDNS}.

The scatter plots of the calculated Hama roughness function on the 35 data samples against the surface statistics involved in calculating the input of neural networks are shown in figure~\ref{fig:dns}b. The generated surfaces in the fine-tuning and pre-training steps show moderately different characteristics. For instance, the surfaces generated in the fine-tuning step show higher limits of the effective slopes (up to $\sim 0.8$) compared to those in the pre-training step (up to $\sim 0.35$). However, note that despite the larger effective slope, the surfaces are not in the regime where drag tend to decrease with increasing effective slope~\citep{flack2014roughness}. The 35 data samples were split into 6, 4, and 25 data samples for training, validation, and test, respectively. More than 70\% of the data samples were used for testing, i.e.,~not used during the fine-tuning step, to fairly evaluate the generalization ability of the developed neural networks. Note that a total of 10 data samples were used in the fine-tuning step, and these fine-tuning data samples are completely separated from the test data samples. The data used for fine-tuning and testing the neural networks is provided in table~\ref{tab:data}. The surface statistics and the Hama roughness function in the test dataset are distributed widely in the total dataset as shown by the $\times$ markers in figure~\ref{fig:dns}b.

To reduce the uncertainty arising from splitting a very small dataset into training and validation sets, we adopted an approach based on averaging many neural networks.
More specifically, we employed 210 neural networks, $\{NN^{1}, NN^{2}, ... , NN^{210}\}$, to learn from the 210 different data combinations derived by selecting 6 training and 4 validation data samples from the 10 fine-tuning data samples
(${10 \choose 6} ={10 \choose 4} =10!/(6!\,4!)=210$). 
The weights and biases of these 210 neural networks ($\{W^{1}_{j,k},W^{2}_{j,k},...,W^{210}_{j,k}\}$ and $\{b^{1}_{k},b^{2}_{k},.., b^{210}_{k}\}$) were initialized by the pre-trained weights and biases, $W^{pre}_{j,k}$ and $b^{pre}_{k}$. Similar to the pre-training step, the weights and biases of each neural network $NN^{i}$ were updated by the Adam optimizer that minimizes a loss for fine tuning $L^i_{f}$,
\begin{equation}
    L^i_{f} = \left (\Delta U^{+} - \Delta \widetilde{U}_{i}^{+}\right )^{2} + \frac{1}{N_w} \sum_{j,k} \mathrm{tr}\left\{(W^{i}_{j,k})^{T}(W^{i}_{j,k})\right\}.
\end{equation}
Here, $\Delta U^{+}$ is the ground truth, and $\Delta \widetilde{U}_{i}^{+}$ is the roughness function predicted by $NN^{i}$.
After a sufficient amount of updates, the networks with the smallest fine-tuning loss on the validation data samples were chosen to predict the roughness function $\Delta U^{+}$ on test data (section~\ref{sec:Result}).

\section{Prediction of roughness functions}\label{sec:Result}
We compare the predicted Hama roughness functions on the 25 test data samples that were not included in the fine-tuning step. The roughness functions are calculated by the following prediction methods: (1) neural networks trained with transfer learning (NNTF), (2) neural networks trained without transfer learning (NN), and (3) empirical correlations (EMP). Neural networks in NN were trained without the pre-training step, i.e., the weights and biases of the networks were randomly initialized instead of being initialized by $W^{pre}_{j,k}$ and $b^{pre}_{k}$. Our focus here is on the comparison between NNTF and NN to demonstrate the effects of transfer learning. However, we also include EMP predictions in the comparison since these methods are widespread in the literature.

The performance of neural networks can vary greatly depending on the selection of data for training and validation. This is because it is generally hard to form a distribution covering the entire feature space from a limited number of data samples. By ensemble averaging the predictions from neural networks learned from different data combinations, we can partly alleviate this problem.
The ensemble predictions from NNTF and NN are thus obtained by averaging the prediction values of the 210 neural networks as,
\begin{equation}
    \Delta \widetilde{U}^{+} = \sum_{i=1}^{210} \Delta \widetilde{U}_{i}^{+}/210.
    \label{eq:u_ensemble}
\end{equation}
Similarly, we use ensemble predictions of the empirical models $\{$\ref{EMP1}, \ref{EMP2}, \ref{EMP3}$\}$ in EMP as,
\begin{equation}
    \Delta \widetilde{U}_{emp}^{+} = \sum_{i=1}^{3} \Delta \widetilde{U}_{i,emp}^{+}/3,
\end{equation}
where $\Delta \widetilde{U}_{i, emp}^{+}$ is the Hama roughness function predicted by the $i$-th empirical model. 

\begin{figure}
  \centering
  \includegraphics[width = 0.5 
  \linewidth]{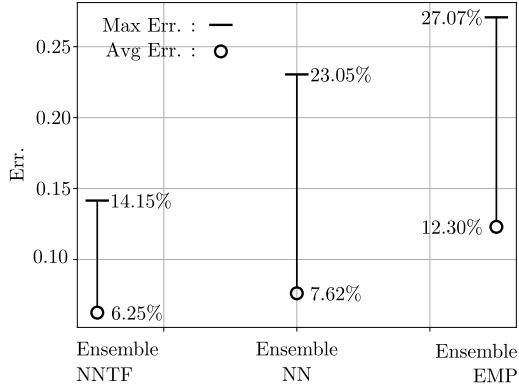}
  \caption{Average and maximum errors from the ensemble predictions performed by NNTF, NN, and EMP.}
\label{fig:ensemble}
\end{figure}

Figure~\ref{fig:ensemble} shows the maximum and average errors calculated from the ensemble predictions by NNTF, NN, and EMP. The predictions from NN exhibit 7.62\% and 23.05\% of average and maximum errors, respectively, while the corresponding error associated with the predictions from NNTF are 6.25\% and 14.15\%. Note that both NN and NNTF are performing better than EMP here. 
The most significant advantage of using transfer learning is in the reduction of the maximum error; nearly ten percent of decrease is achieved by using NNTF instead of NN. As we will discuss more quantitatively in the next section, this error reduction demonstrates the capability of NNTF in learning a better generalized mapping from a small amount of DNS data.

\begin{figure}
  \centerline{\includegraphics[width = 1.0 \linewidth]{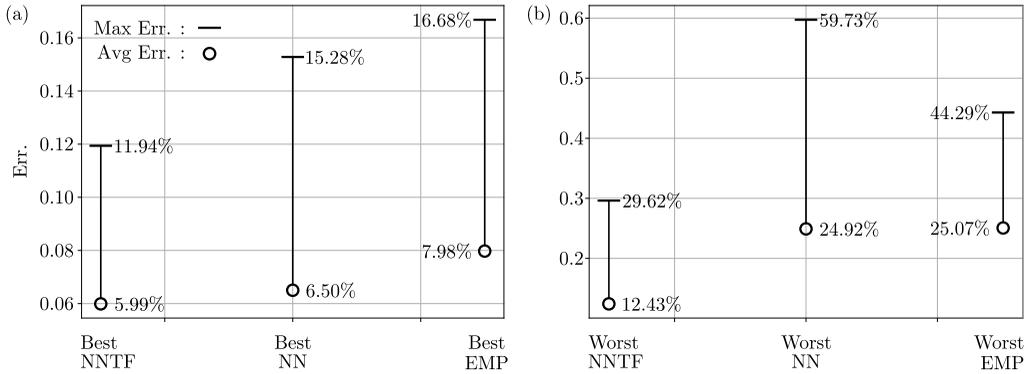}}
  \caption{Average and maximum errors from the (a) best and (b) worst performing neural networks and empirical correlations.}
\label{fig:bestandworst}
\end{figure}

To further investigate transfer learning, the best and worst performing neural networks out of the 210 networks in NNTF and NN were extracted and compared in figure~\ref{fig:bestandworst}. The errors from the best performing networks in NNTF and NN (figure~\ref{fig:bestandworst}a) show a similar trend as the ensemble predictions (figure~\ref{fig:ensemble}); both average and maximum errors are clearly reduced by using transfer learning. We also note that the ensemble predictions from NNTF (figure~\ref{fig:ensemble}) marginally perform better compared to both the best performing neural network in NN and empirical correlation in EMP (figure~\ref{fig:bestandworst}a). 

The advantage of using transfer learning is more clearly demonstrated in the worst performing case (figure~\ref{fig:bestandworst}b). The maximum error of the worst performing neural network in NN is nearly 60\%, showing that this network has failed to provide a reliable prediction. On the other hand, the network in NNTF exhibits significantly smaller errors, indicating that transfer learning is helping the network to provide a safer prediction even in the worst scenario. Therefore, these results show that transfer learning enables the networks to learn a more generalized mapping from limited DNS data by adapting an `approximate’ knowledge from empirical correlations. It is also important to emphasize that ensemble predictions should be used in practice to reduce the uncertainty arising from a limited pool of data as can be seen from the wide range of errors that occur with the best and worst performing neural networks in figure~\ref{fig:bestandworst}. 

\section{Analysis of transfer learning and approximate knowledge}\label{sec:discussion}
In the previous section, we demonstrated that transfer learning of empirical correlations improves the learning performances of neural networks. In this section, we analyze the effects of transfer learning and why `approximate knowledge' in the empirical correlations can help neural networks.

\subsection{Effects of transfer learning on weight generalization}\label{sec:weights}
In this section, we investigate how transfer learning improves the generalization ability by characterizing the learning weights of the networks. Let $w^{i}_{l}$, where $l=1,2,...,N_{w}$, be the vectorized elements of the weight matrices $W^{i}_{j,k}$ in $NN^{i}$ ($N_{w}=488$ in this study). Then, we define the deviation of weights $w^{i}_{l}$ from different $NN^{i}$ as
\begin{equation}
  \sigma_{l} = \frac{\sqrt{\sum_{i=1}^{210}\left(w^{i}_{l} - \mu_{l}\right)^{2}/210}}{\sum_{i=1}^{210}\left|w^{i}_{l}\right|/210},
\end{equation}
where $\mu_{l}= \sum_{i=1}^{210}w^{i}_{l}/210$. Note that the networks in NNTF were initialized with the pre-trained weights, and the networks in NN were initialized with the same random weights. Therefore, the deviation $\sigma_{l}$ indicates how a weight in a neural network is updated differently for different combinations of training and validation data samples (i.e., different $NN^{i}$). In the ideal case with an infinite amount of data samples, the deviation will approach zero as the network would converge to specific weights representing a unique generalized mapping of the real physics.

\begin{figure}
  \centerline{\includegraphics[width = 0.5 \linewidth]{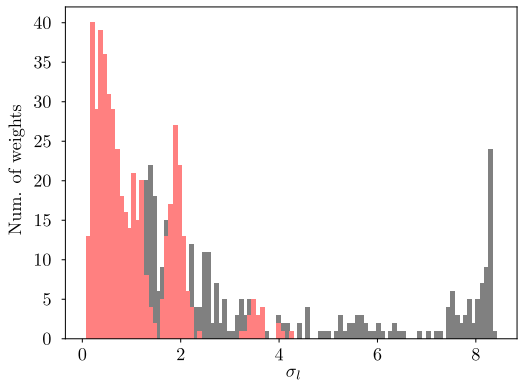}}
  \caption{Distributions of the deviations of weights $\sigma_{l}$ in NNTF (red) and NN (black).}
\label{fig:histo}
\end{figure}

The distributions of the deviation $\sigma_{l}$ calculated from NNTF and NN are compared in figure~\ref{fig:histo}. A number of weights with large $\sigma_{l}$ are observed in NN, indicating that the updates of weights largely deviate depending on the data. This implies that the current data is insufficient for NN to learn a generalized mapping. On the other hand, the deviations are significantly reduced in NNTF, implying that the networks in NNTF are learning a better-generalized mapping compared to NN. This is because the networks in NNTF were initialized with the pre-trained weights containing the `approximate’ knowledge. These weights provide a better initial state for the weight optimization problem in the fine-tuning step.

\subsection{Effects of transfer learning on sensitivities of input surface statistics}\label{sec:sensitivity}
This section investigates how transfer learning affects the neural networks in learning the sensitivities of the input surface statistics.
The sensitivities of input surface statistics in predicting drag are obtained by calculating the derivatives of drag with respect to each input (surface statistical measure).
A similar analysis was employed by ~\citet{kim2020prediction}, for evaluating the influences of flow variables in predicting heat transfer characteristics. 
We define the sensitivity of the input $I_{i}$ in predicting drag $\Delta \widetilde U^{+}$ (defined in equation~\ref{eq:u_ensemble}) as:
\begin{equation}
    S_{i} = \left\langle\left|\frac{\partial \Delta\widetilde U^{+}}{\partial I_{i}}\right|\right\rangle . \label{eq:sensitivity}
\end{equation}
Here, $\langle\cdot\rangle $ is the averaging operation along with the calculated derivatives from the 25 test DNS dataset. The derivative $\frac{\partial \Delta\widetilde U^{+}}{\partial I_{i}}$ can be analytically calculated as neural networks are composed of differentiable function compositions and matrix multiplications. However, more efficiently, it can be done by using an automatic differentiation algorithm. In this study, we used the automatic differentiation algorithm implemented in PyTorch~\citep{paszke2017automatic}.

\begin{figure}
  \centerline{\includegraphics[width = 0.6 \linewidth]{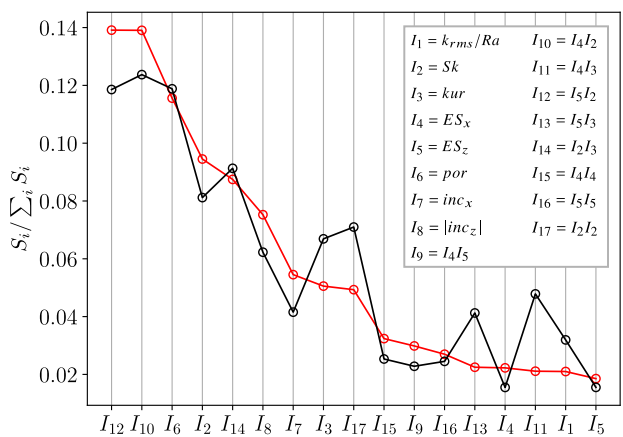}}
  \caption{Sensitivities of the input surface statistics in NNTF (red) and NN (black). The sensitivities are normalized with their total sum ($S_{i}/\sum_{i}S_{i}$). The input surface statistics are sorted in descending order from the largest to smallest sensitive surface statistics in NNTF.}
\label{fig:sen}
\end{figure}

Figure~\ref{fig:sen} shows the sensitivities of the input surface statistics when trained with NNTF and NN. It is observed that for both  neural networks, $I_{12}=ES_{z}\times Sk$,  $I_{10}=ES_{x}\times Sk$ and $I_{6}=por$ have the highest sensitivities. In other words, these are the most important surface measures when it comes to predicting the friction drag for the particular class of rough surfaces contained in the high-fidelity dataset.  
These statistics are not explicitly included in the empirical models \eqref{EMP1}, \eqref{EMP2} and \eqref{EMP3}. Therefore, this indicates that the neural networks are learning the mapping of surface statistics to drag beyond the empirical models. However, the importance of $I_{12}, I_{10}$ and $I_{16}$ is partially  implied by previous studies.
The importance of the product of effective slope and skewness is partly implied in the study of \citet{forooghi2017toward}, as their empirical model (\ref{EMP2}) is nonlinear with respect to skewness and effective slope. The importance of porosity in predicting drag was recently emphasized by \cite{jouybari2021data}.
Although the result shown here is for a particular class of rough surfaces, it demonstrates the necessity of expanding the feature space of surface statistics for drag prediction to more complex nonlinear combinations of the most basic statistical moments. 

To understand what influence transfer learning has on sensitivities, we can identify from figure~\ref{fig:sen}  the input statistics that are ranked differently by NN and NNTF in terms of importance. The surface statistics related to effective slopes and skewness, e.g., $I_{12}=ES_{z}\times Sk$, $I_{10}=ES_{x}\times Sk$, $I_{2}=Sk$, $I_{4}=ES_{x}$, and $I_{5}=ES_{z}$, are found to be more sensitive in NNTF. These higher sensitivities in NNTF are mainly due to the inclusion of the approximate knowledge about the drag dependencies on effective slope and skewness in the empirical correlations. Conversely, the sensitivity of the input defined as the square of skewness ($I_{17}=Sk \times Sk$) is smaller in NNTF compared to NN. This indicates that NNTF learns that the square of skewness compromises the information about the sign of skewness, which is known to be important for drag prediction~\citep{jelly2018reynolds}. In addition, NNTF reduces the sensitivities of statistics related to kurtosis ($I_{3}=kur$, $I_{13}=ES_{x}\times kur$, and $I_{11}=ES_{z}\times kur$). As kurtosis measures the outliers~\citep{decarlo1997meaning}, the reduction indicates that the effects of the outlier roughness heights are relaxed in NNTF.

\subsection{Approximate knowledge in the empirical correlations}\label{sec:approximate}
\begin{figure}
  \centerline{\includegraphics[width = 1.0 \linewidth]{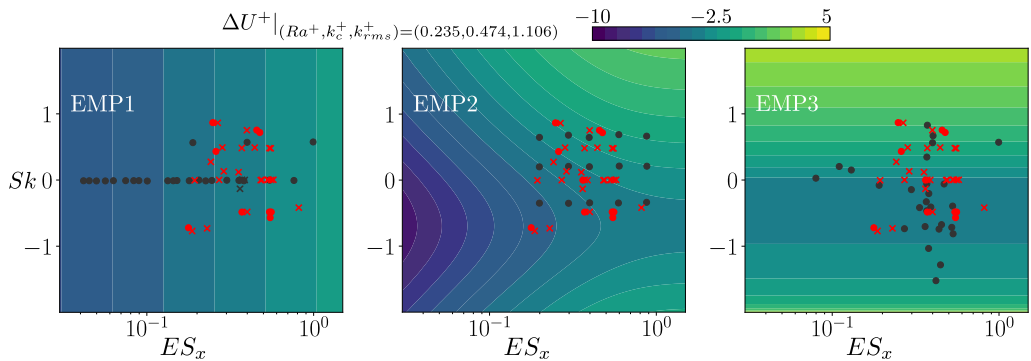}}
  \caption{Contours of the roughness function calculated from \ref{EMP1}, \ref{EMP2}, \ref{EMP3}. The roughness heights $(Ra^{+},k_{c}^{+},k_{rms}^{+})=(0.234,0.474,1.106)$) are used for calculating the empirical correlations. The red {\color{red}$\times$} and {\color{red}$\circ$} markers indicate the test and fine-tuning DNS data samples, respectively. The black $\circ$ markers indicate the fitting data samples used for deriving the empirical datasets (data extracted from~\cite{chung2021predicting}).}
\label{fig:contour}
\end{figure}

As briefly introduced in $\S$~\ref{sec:pre-train}, the empirical correlations contain different types of approximate knowledge. This is because the empirical correlations were fitted from different types of surfaces~\citep{chung2021predicting}; \ref{EMP1} was derived from data of a pipe with sinusoidal irregular roughness; \ref{EMP2} was developed from data of rough surfaces with different arrangements and size distributions of roughness elements and finally; \ref{EMP3} was constructed from a wide range of realistic rough surfaces, such as gravels and commercial steel pipes. As a consequence, each of the relations \ref{EMP1}, \ref{EMP2} and \ref{EMP3} depends on a different measure of roughness height ($Ra^{+}$, $k_{c}^{+}$, or $k_{rms}^{+}$) and different combinations of surface statistics ($Sk$ and/or $ES_{x}$). 
Figure~\ref{fig:contour} shows contour levels of the drag predicted by  \eqref{EMP1}, \eqref{EMP2} and \eqref{EMP3} in a map spanned by skewness $Sk$ and effective slope $ES_x$. Note that the roughness heights for \ref{EMP1} to \ref{EMP3} ($Ra^{+}$, $k_{c}^{+}$, and $k_{rms}^{+}$) are chosen by their respective values that produce $\Delta U^{+}=-3.5$ at a common location ($ES_{x} = 0.3$ and $Sk = 0$) as in \cite{chung2021predicting}.  

The correlation \ref{EMP1} contains the approximate knowledge that drag tends to increase with increasing effective slope, while it does not contain any dependency of drag with respect to skewness. This is because \ref{EMP1} was derived from fitting data with a narrow regime of skewness as shown in the black $\circ$ markers in EMP1 of figure~\ref{fig:contour}. Oppositely, the correlation \ref{EMP3} contains the knowledge that drag tends to increase with increasing skewness, without the dependency on effective slope. This is because the majority of the fitting data for \ref{EMP3} (black $\circ$ markers in EMP3 of figure~\ref{fig:contour}) lies on an insensitive regime of effective slope -- between the sparse regime ($ES<0.3-0.6$) and dense regime ($ES>0.4-3.0$) -- with relatively small effects on drag~\citep{jimenez2004turbulent}. \ref{EMP2} shows a nonlinear dependence of drag on the skewness and the effective slope. 
As a result, this model includes approximate knowledge about the drag dependency on the product of skewness and effective slope as also discussed in \S\ref{sec:sensitivity}.

As the empirical correlations are derived by fitting data from a particular set of surfaces, a naive extrapolation of them for predicting drag of other set of rough surfaces may lead to large errors. For example, if the correlations are directly applied to predict drag on the current test surfaces of the DNS dataset (red {\color{red}$\times$} markers in figure~\ref{fig:contour}), the average and maximum errors of \ref{EMP1}, \ref{EMP2}, and \ref{EMP3} are ($17\%$, $29\%$), ($8\%$, $17\%$) and ($25\%$, $44\%$), respectively. Note that the errors from \ref{EMP2} are the smallest due to the similar surface spaces between the current DNS data and the fitting data. 
Interestingly, large errors of the empirical models do not necessarily lead to inaccurate fine-tuned neural networks. To demonstrate this, we  fine-tuned neural networks pre-trained only with one empirical correlation. We found that the average and maximum errors of the ensemble networks fine-tuned from \ref{EMP1}, \ref{EMP2}, and \ref{EMP3}, were ($6\%$, $18\%$), ($6\%$, $17\%$), and ($7\%$, $14\%$), respectively. Despite the high errors of the empirical correlations of \ref{EMP1} and \ref{EMP3}, their resulting fine-tuned networks are as effective as the fine-tuned network from \ref{EMP2}. This indicates that an empirical model for pre-training does not necessarily need to be accurate, it merely needs to provide an approximate knowledge of the dependencies of on surface statistics.
Moreover, NNTF, which is pre-trained with all of the approximate knowledge provided by the three empirical models, achieves a more favorable overall performance (figure~\ref{fig:ensemble}), compared to the fine-tuned networks pre-trained with any single EMP. This is because the knowledge contained in each empirical model contributes with information of drag dependencies. 
This can be qualitatively shown by visualizing knowledge domains. 

\begin{figure}
  \centerline{\includegraphics[width = 1.0 \linewidth]{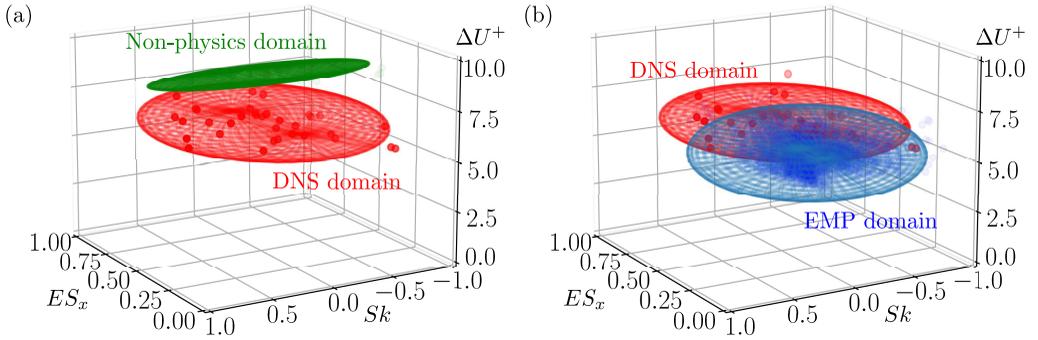}}
  \caption{Visualization of the knowledge domains of the DNS domain (red), EMP domain (blue), and Non-physics domain (green). Comparisons between the (a) DNS and Non-physics domains and the (b) DNS and EMP domains are shown.}
\label{fig:knowledge}
\end{figure}

Figure~\ref{fig:knowledge} compares the knowledge domains of high-fidelity physics (DNS domain), approximate knowledge in the empirical models (EMP domain), and non-physical knowledge from the randomly initialized neural networks (Non-physics domain).
To visualize the knowledge domains, we extracted principal axes of the data samples composed of $(ES_{x}, Sk, \Delta U^{+})$ in each domain. Note that $\Delta U^{+}$ in the DNS, EMP, and Non-physics domains are obtained from simulations, empirical models, and randomly initialized neural networks, respectively. The DNS domain is composed of rough surfaces in the DNS dataset, while the EMP and Non-physics domains are composed of rough surfaces in the empirical dataset.
After computing the principal axes in each domain, ellipsoids that approximately bound the DNS, EMP, and Non-physics domains along their principal axes are visualized. As shown in figure~\ref{fig:knowledge}a, no particular correlation (or rather alignment) can be found between the Non-physics domain and the DNS domain. On the other hand, an alignment between the EMP domain and the DNS domain is observed in figure~\ref{fig:knowledge}b. Therefore, it is expected that neural networks with approximate knowledge from the EMP domain can more easily adapt to the DNS domain compared to those without any physical knowledge.

We also quantified the alignment between the knowledge domains by calculating the angles between the principal axes. The angles between the principal axes in the EMP and DNS domains are computed as ($16.5^{\circ}$, $18.1^{\circ}$, $8.9^{\circ}$), while the angles between the principal axes in the Non-physics and DNS domains are computed as ($120.9^{\circ}$, $119.9^{\circ}$, $12.3^{\circ}$). 
Accordingly, the aligned `approximate knowledge' from EMPs assists the networks to adapt to the DNS domain; thus, a better performance of neural networks can be achieved by transfer learning of empirical correlations. 

\section{Conclusions}\label{sec:conclusion}
We have developed a transfer learning framework to learn the Hama roughness function from a limited number of DNS data samples. The framework is composed of two steps: (1) pre-training step and (2) fine-tuning step. In the pre-training step, neural networks learn an `approximate' knowledge from empirical correlations. In the fine-tuning step, the networks are fine-tuned using a small DNS dataset. Neural networks trained with transfer learning have shown a significant improvement in predicting the roughness function on test data samples not included in the fine-tuning step. This is because the `approximate' knowledge in empirical correlations is aligned to high fidelity physics, which assists neural networks to learn better-generalized weights. In addition, a sensitivity analysis showed that the neural networks with transfer-learning were clearly emphasizing the importance of certain input surface statistics (effective slopes, skewness and porosity) in predicting drag. These extracted statistics are in general  good agreement with what has been reported by other investigations, but also highlight that nonlinear functions of several surfaces statistics could provide high correlation with drag.

We have shown that the prediction performance is enhanced when the `approximate' knowledge of the empirical dataset is well aligned to the high-fidelity dataset. Therefore, it is advantageous to employ empirical correlations developed for classes of rough surfaces that are similar to one's high-fidelity dataset. 
The current NNTF, trained with irregular homogeneous rough surfaces, would not be effective when predicting drag on regular rough surfaces (e.g.,~cuboids, bars, etc.) or inhomogeneous rough surfaces. Similarly, predicting values of roughness function beyond training is also expected to not be effective due to the general limitations of data-driven methods in extrapolating.

To further increase the generalization ability of neural networks, ultimately, the dataset has to be expanded. A large database would also help to study classes of rough surfaces with undiscovered empirical correlations, where pre-training is currently not possible. Accordingly, a collective community effort is needed to construct and train Big Data of flows over many classes of rough surfaces in different flow conditions. One step in this direction, is the online database \href{http://roughnessdatabase.org/}{Roughness database}~\citep{roughnessdatabase} currently under construction. Moreover, new primary surface statistics that strongly affect drag may be discovered in the future by extending the sensitivity analysis in \S\ref{sec:sensitivity} with a larger surface parameter space and Big Data. 
Also, expansion of the network structure would enable one to fully leverage information inside large databases. The structure of the neural network in this study is very simple, as our focus has been to show that transfer learning can improve the performance of a given network. Thus, this network structure should not be considered as the optimal model for predicting drag. In addition to expanding the database, further optimizing the neural networks will also improve the performance for drag prediction.

Finally, while we have considered transport of momentum (i.e., drag coefficient), empirical relations are widespread also for characterizing transport of energy (Nusselt number) and mass (Sherwood number). Therefore, the proposed framework can also be applied to other engineering applications where only a limited amount of high-fidelity data is available, but a significant amount of knowledge has been accumulated.

\section*{Acknowledgements}{This work was supported by the Swedish Energy Agency under Grant number 51554-1, Swedish Foundation for Strategic Research (FFL15-001), and Friedrich und Elisabeth Boysen-Foundation (BOY-151). SL and SB would like to thank Dr. Oscar Beijbom for the helpful discussions regarding the neural networks.}

\section*{Declaration of interests}{The authors report no conflict of interest.}

\appendix

\section{Effects of a bounded roughness function in the datasets}\label{appBU}
Here, we provide an example of training the neural networks with bounded $\Delta U^{+}$. If the bound of $\Delta U^{+}$ in one’s high-fidelity dataset is known, the transfer learning performance can be improved by accordingly bounding the $\Delta U^{+}$ for pre-training. In this example, we used the bound of $\Delta U^{+}>6$, as the surfaces in the current DNS dataset asymptotically reach the fully rough regime when $\Delta U^{+} \approx 6$~\citep{jiasheng2021dns}. We excluded the data samples with $\Delta U^{+} < 6$ from the 35 DNS data samples (figure~\ref{fig:dns}). As a result, a total of 29 DNS data samples are used in this section. The 29 data samples were split into 5, 3, and 21 data samples for training, validation, and testing. Note that more than 70\% of the data samples were used for testing as in~\S\ref{sec:Result}. The $\Delta U^{+}$ of the empirical dataset is bounded to $\Delta U^{+}>6$ as the DNS dataset. After pre-training neural networks with the bounded $\Delta U^{+}$, we fine-tuned a total of 56 neural networks, which learn from the 56 different combinations of 5 training and 3 validation data samples out of the total 8 fine-tuning data samples (${8 \choose 5} ={8 \choose 3} =8!/(5!\,3!)=56$).

\begin{figure}
  \centerline{\includegraphics[width = 0.5 \linewidth]{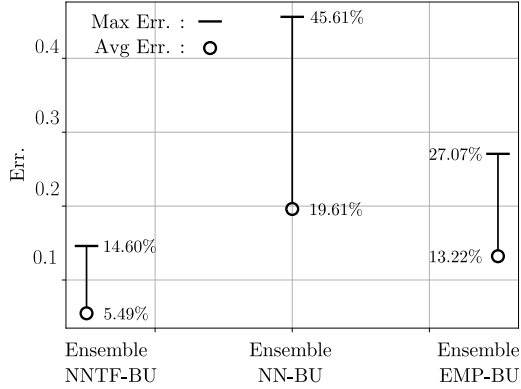}}
  \caption{Average and maximum errors from the ensemble predictions performed by NNTF-BU, NN-BU, and EMP-BU.}
\label{fig:BU}
\end{figure}
The average and maximum errors on the 21 test data samples predicted by the ensembles of (1) NNTF with bounded $\Delta U^{+}$ (NNTF-BU), (2) NN with bounded $\Delta U^{+}$ (NN-BU), and (3) EMP with bounded $\Delta U^{+}$ (EMP-BU) were ($5.49\%$ and $14.60\%$), ($19.61\%$ and $45.61\%$), and ($13.22\%$ and $27.07\%$), respectively (figure~\ref{fig:BU}). 
The errors from NN-BU were notably increased compared to the errors from NN ($7.62\%$ and $23.05\%$, figure~\ref{fig:ensemble}). This is because the number of data samples used for training NN-BU has decreased by $\sim 20\%$ compared to the number used for training NN. However, despite the decrease of the number of fine-tuning data samples, NNTF-BU shows similar errors compared to NNTF ($6.25\%$ and $14.15\%$, figure~\ref{fig:ensemble}). Accordingly, NNTF-BU achieved nearly 15\% and 30\% decrease in error percentages compared to NN-BU, which is more significant than those achieved from NNTF (figure~\ref{fig:ensemble}). Therefore, the transfer learning performance can be improved when the range of $\Delta U^{+}$ of surfaces in one’s high-fidelity dataset is \textit{a priori} known. It is also worth mentioning that the errors from EMP-BU are similar to those from EMP ($12.30\%$ and $27.07\%$, figure~\ref{fig:ensemble}), which indicates that the employed empirical models are reasonably valid even for $\Delta U^{+}$ slightly under 6.

\section{Grid resolution and domain size}\label{appDNS}
\begin{figure}
  \centerline{\includegraphics[width = .7 \linewidth]{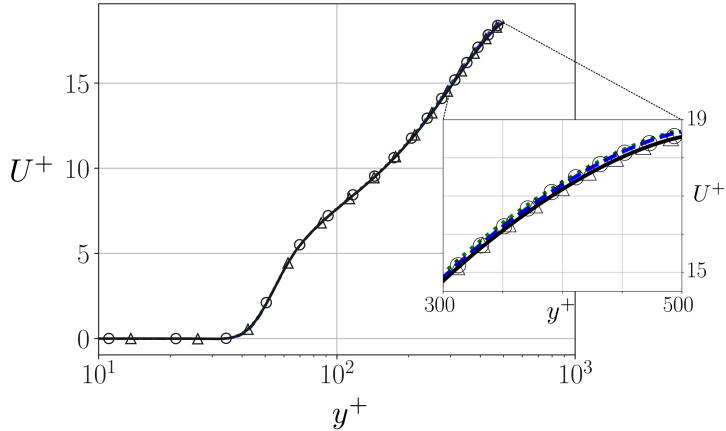}}
  \caption{Profiles of $\Delta U^{+}$ calculated from Case 1 (green dotted line), Case 2 (blue dashed line), Case 3 (black solid line), Case4 ($\circ$), and Case 5 ($\bigtriangleup$).}
\label{fig:grid}
\end{figure}

To determine the appropriate mesh resolution in our simulations, a grid independence test was conducted. We conducted the test for the rough surface with the smallest Taylor microscale ($\lambda_{T}$) along with the 35 DNS data samples. The definition of the Taylor microscale proposed by \cite{yuan2014estimation} is adopted in the current study. We studied the grid independence using five different resolutions in a minimal channel domain $(2.0H \times 2.0H \times 0.4H)$: Case 1, $(192 \times 401 \times 36)$; Case 2, $(256 \times 361 \times 48)$; Case 3, $(256 \times 401 \times 48)$; Case 4, $(256 \times 451 \times 48)$; Case 5, $(300 \times 401 \times 60)$. The calculated mean velocity ($U^{+}$) profiles are shown in figure~\ref{fig:grid}. The profiles from the five different cases are found to be nearly identical in both the inner and outer layers. Thus, in the current study, we used the grid resolution of Case 3. This resolution corresponds to $\Delta^{+}<5$ in both streamwise and spanwise directions. In addition, with this resolution, the grid sizes were at least four times smaller than the Taylor microscales of all surfaces in both streamwise and spanwise directions $(\lambda_{T}/\Delta>4.5)$, satisfying the grid resolution constraint introduced in~\citet{jouybari2021data}.

The sensitivity of the domain size was also studied. We performed additional 12 full channel DNS to inspect the feasibility of using minimal channel DNS in the current study. For a given channel half height $H$, the domain and grid sizes for the full channel simulations were, respectively, $(8H \times 2H \times 4H)$ and $(900 \times 401 \times 480)$, in the streamwise, wall-normal, and spanwise directions. The smallest domain and grid sizes for the minimal channel simulations were $(2.0H \times 2.0H \times 0.4H)$ and $(256 \times 401 \times 48)$, respectively. The surface statistics of the rough surfaces were reasonably well converged in both domains. For instance, the maximum differences of $k_{rms}^{+}$, $ES$, and $Sk$ between the full and minimal channel domains for the rough surfaces were $0.038$, $0.025$, and $0.006$, respectively. The resulting average and maximum errors between $\Delta U^{+}$ calculated from the full and minimal channel domains were $1.5\%$ and $4.6\%$, respectively.
As these errors are notably smaller than those from the predictions of neural networks (\S\ref{sec:Result} and Appendix~\ref{appBU}), we found the current minimal channel approach sufficient for our purposes. The full details of the surfaces and simulations are available in~\cite{jiasheng2021dns}.

\bibliographystyle{jfm}
\bibliography{SLee}

\end{document}